\documentclass[fleqn,twoside]{article}
\usepackage{espcrc2,psfig}

\usepackage{graphicx}
\newcommand{\AmS}{{\protect\the\textfont2
  A\kern-.1667em\lower.5ex\hbox{M}\kern-.125emS}}

\title{The continuum limit of perturbative coefficients 
calculated with a large field
cutoff}
\author{Li Li\address[MCSD]{Department of Physics and Astronomy, \\
The University of Iowa, \\
Iowa City, Iowa 52242, USA}
       and
        Y. Meurice\addressmark[MCSD] \thanks{This 
research was supported in part by the Department of Energy
under Contract No. FG02-91ER40664.
We thank the Department of Phys. and Astr. and the ITS for making possible
the use of Linux clusters. Y. M. was partially supported by a 
Faculty Scholar Award at The University of Iowa and a residential
appointment at the Obermann Center for Advanced Studies at the
University of Iowa, while these proceedings were written.
}}
\begin{document}
\begin{abstract}
We report MC calculations of perturbative coefficients for lattice
scalar field theory in dimensions 1, 2 and 3, where the large field
contributions are cutoff. This produces {\it converging} (instead of 
asymptotic) perturbative
series. We discuss the statistical errors and the lattice effects and
show that accurate calculations are possible even in a crossover
region where no approximation works. We show that the field cutoff is
also a UV regulator. We point out the relevance for QCD questions discussed
by Tomboulis and Trottier at this conference.
\vspace{1pc}
\end{abstract}
\maketitle

A central problem in quantum field theory is the presence of different
behaviors at different scales. 
This is clearly the case for QCD \cite{tomboulis} 
where the short distance behavior can be described using perturbation 
theory, while confinement and other non-perturbative phenomena
appear at large distance. A 
somehow similar situation is encountered in scalar and spin 
models. We limit here the discussion to the symmetric phase where
a trivial high-temperature (strong coupling) fixed point is present.
Perturation theory generically fails to provide a proper description of
the RG flows near the HT fixed point. From this point of view, 
the fact that the perturbative series are asymptotic is not
surprising. 

Asymptotic series can be a serious problem
for problems where the strong interactions correction are taken into
account. A good example is the hadronic width of the $Z$, where the
third order corrections\cite{larin94} are about two-third of the second order
corrections and larger than the experimental error bar for the
combined LEP experiments. Perturbative methods also play an important
role in the improvement methods\cite{trottier} that were used 
to obtain unprecedented accuracy for quantities that can be compared 
with experiments\cite{davies}. Given these recent successes, the
standards of accuracy are raised and 
it is likely that the lack of 
convergence of QCD series will need to be addressed in the near future.

A simple way\cite{pernice98,convpert} 
to convert an asymptotic series into a convergent one consists in 
introducing a large field cutoff.
It has been shown with three 
$\phi^4$ problems \cite{convpert} that 
the modified series {\it converge}
toward values which are exponentially close to the exact ones.
At a fixed order, it is possible to chose
the field cutoff in order to optimize the accuracy\cite{optimal}.
The three examples discussed in Ref. \cite{convpert} 
can be solved accurately with 
numerical methods \cite{anh,gam}. However, for generic problems, this
kind of 
calculation can be quite difficult, especially in a crossover region 
(illustrated in Fig. 3) where neither semi-classical methods or 
HT expansion are available. 
This difficulty reflects the fact that, in general, the 
interpolation between RG fixed
points is a difficult non-linear problem 
that has only be solved for simplified models (see
\cite{scaling} for an example).

When no other methods are available, one has to resort to the 
MC method to perform calculations of the modified 
perturbative coefficients. In the following,
we consider lattice scalar models  
with nearest neighbor interactions and one component at each site, 
in 1, 2 or
3 dimensions. For a lattice with $N_s$ sites, we will consider the 
following quantities relevant for perturbation theory:
\begin{eqnarray}
A_0&=&<\sum_x \phi^2_x>/N_s\\ 
A_1&=&<\sum_x \phi^4_x>/N_s \\
A_2&=&(<(\sum_x \phi^4_x)^2>-<\sum_x \phi^2_x>^2)/N_s\ .
\end{eqnarray}
The average $<>$ correspond to a Gaussian model where the integration
at each site goes from $-\phi_{max}$ to $\phi_{max}$.
In $D=1$, the continuum values for $\phi_{max}=\infty$ are  
$A_0=1/2$, $A_1= 3/4$ and $A_2=21/8$. For $D$=2 or 3, $A_0$ is UV 
divergent.
Two kinds of 
errors that should be taken into account: the statistical errors and
the errors due to the finite lattice spacing. 

\begin{figure}[htb]
%\centerline{\psfig{figure=fig1-1.eps,width=2.2in}}
%\centerline{\psfig{figure=fig1-2.eps,width=2.2in}}
\vskip-5pt
\centerline{\psfig{figure=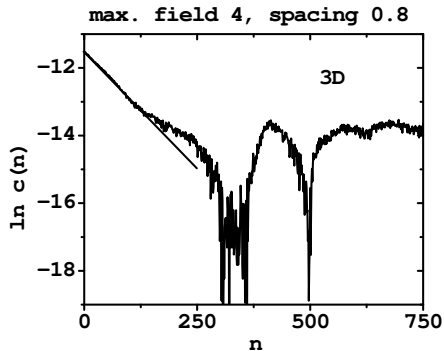,width=2.6in}}
\vskip-25pt
\caption{Correlation functions for 
the observable $A_0$ in $D=3$ as a function of the separation $n$
in the configuration number.}
\end{figure}

The
statistical errors decrease as the inverse square root of the number of
decorrelated configurations.
The correlation times are estimated from the exponential decay of the 
correlations. If we denote by $O_i$ the value of the observable 
$O$ in a configuration $i$, the correlations $C(n)$ are subtracted
averages of
$O_iO_{i+n}$. Fig. 1 illustrates this exponential decay in an example.
Alternatively, we can use Ref. \cite{mandula} to estimate 
the statistical error for correlated data.
%\begin{figure}[htb]
%\centerline{\psfig{figure=a1d1.eps,width=2.2in}}
%\centerline{\psfig{figure=d1f4-sub.eps,width=2.5in}}
%\caption{Approach of the zero lattice spacing for the first
%coefficient of the ground state of the 
%anharmonic oscillator at various field cuts. The
%infinite cut value is 3/4.}
%\end{figure}
\begin{figure}[htb]
\vskip-15pt
\centerline{\psfig{figure=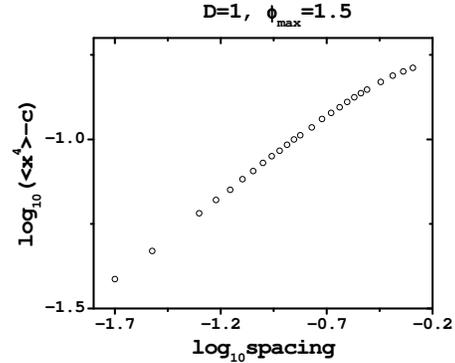,width=2.6in}}
\vskip-15pt
\caption{Illustration of the power corrections to $A_1$ in $D=1$. 
We have subtracted 
the estimated continuum value $c$.}
\end{figure}

Second, we need to take into account the lattice effects. 
For small enough lattice spacing $a$, a simple power behavior
$c+a^\alpha$ is observed, $c$ being the continuum value that we are trying
to calculate. The power behavior can be extracted from
the statistical fluctuations (which increase when $a$ becomes small!)
for an intermediate range of lattice spacing, using a nonlinear fit.
This is illustrated in Fig. 2. More detail will be provided in \cite{lili}.

The comparison between the MC results obtained with the method 
described above and the accurate numerical
results obtained with methods inspired by Ref. \cite{anh} is shown
below in the case $D=1$ . The MC estimates for $A_1$ for various
$\phi_{max}$ are compared with accurate values in the table below
\vskip15pt
\centerline{\begin{tabular}{|c|c|c|}
\hline
$\phi_{max}$ & MC  & Num.\\
\hline
3 & 0.738(4) & 0.7333 \\
2.5 & 0.646(12)& 0.6482 \\
2 & 0.422(9) & 0.4315 \\
1.5 & 0.178(2)  & 0.1807 \\
1 & 0.03987(3)  & 0.03995 \\
\hline
\end{tabular}}
\vskip15pt
The results for $A_1$ and $A_2$ are shown in Fig. 3.
The MC estimates for $A_2$ require a subtraction and are less
accurate. We expect that 
better results can be obtained by improving
the statistics and refining the nonlinear fit method.
%\vskip5pt
\begin{figure}[htb]
\vskip-5pt
\centerline{\psfig{figure=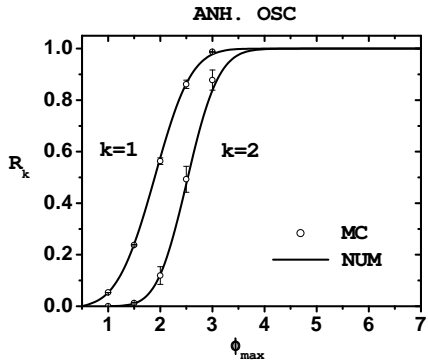,width=2.6in}}
\vskip-15pt
\caption{MC (circles) versus accurate numerical results (continuous
line) for $A_1$ (left) and $A_2$ (right). All the values have been  
divided by their respective infinite cut limits 3/4 and 21/8. }
\vskip-10pt
\end{figure}

It is possible to calculate approximately the values at the bottom
(top) of the curves of Fig. 3 using a HT expansion
(semi-classical methods) and reliable interpolation methods are being
developed to apprach the crossover region.

$A_0$ diverges like Ln($a$) in $D=2$ and $1/a$ in $D=3$, when the
lattice spacing $a\rightarrow 0$. However, the field cut takes care of
this divergence. This is illustrated in Fig. 4 for $D=2$. This
regularization
has a simple explanation\cite{kobe}: since the average (1) provides a 
positive measure, we can obtain a bound by replacing $\phi_x^2$ by its
maximal value. This yields the bound $A_0 \leq \phi_{max}^2$.
The field cut can be thought as produced by an interaction of the 
form $(\phi/\phi_{max})^s$ in the large $s$ limit. Consequently, this
should not affect the universal features of the model and it should be
possible to calculate the critical exponents using modified
perturbative methods. We are 
planning to extend these methods to gauge theories.

\begin{figure}
%\centerline{\psfig{figure=a1d1.eps,width=2.6in}}
\vskip-6pt
\centerline{\psfig{figure=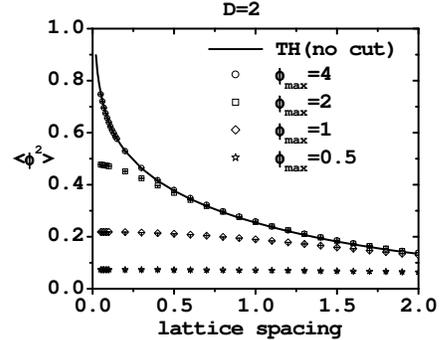,width=2.5in}}
\vskip-16pt
%\centerline{\psfig{figure=c:/talks/lat03/p3/f2d3.eps,width=2.5in}}
\caption{$A_0$ for $D$ = 2
as a function of the lattice spacing for various field cuts. 
The continuous line is the (calculable) lattice result (without cut).}
\end{figure}

\end{document}